\documentclass[10pt, conference, compsocconf]{IEEEtran}   
\usepackage{graphicx}
\usepackage{amsmath}
\usepackage{booktabs}
\usepackage{amsfonts}
\usepackage{multirow}
\usepackage[ruled,vlined]{algorithm2e}
\usepackage{algpseudocode}
\usepackage{subcaption}
\usepackage{indentfirst}
\usepackage{booktabs, enumitem}
\usepackage[export]{adjustbox}
\usepackage{xcolor}

\newcommand{\flop}{\mathrm{flop}}
\newcommand{\nnz}{\textit{nnz}}
\newcommand{\cf}{\textit{cf}}
\newcommand{\erdosrenyi}{Erd\H os-R\'{e}nyi}

\newcommand{\mA}{\mathbf{A}} 
\newcommand{\mB}{\mathbf{B}}

\newcommand{\mC}{\mathbf{C}}

\title{\LARGE Bandwidth-Optimized Parallel Algorithms for Sparse Matrix-Matrix Multiplication using Propagation Blocking}

\author{Zhixiang Gu$^*$, Jose Moreira$^\dagger$, David Edelsohn$^\dagger$, Ariful~Azad$^*$,  \\
  {\small E-mail: zg3@iu.edu, jmoreira@us.ibm.com, edelsohn@us.ibm.com,  azad@iu.edu }\\
  $^*$Indiana University Bloomington, 
  $^\dagger$IBM Corporation\\
  
}

\date{}

\begin{document}

\maketitle

\begin{abstract}
 Sparse matrix-matrix multiplication (SpGEMM) is a widely used kernel in various graph, scientific computing and machine learning algorithms.  
 It is well known that SpGEMM is a memory-bound operation, and its peak performance is expected to be bound by the memory bandwidth.
 Yet, existing algorithms fail to saturate the memory bandwidth, resulting in suboptimal performance under the Roofline model. 
 In this paper we characterize existing SpGEMM algorithms based on their memory access patterns and develop practical lower and upper bounds for SpGEMM performance. 
 We then develop an SpGEMM algorithm based on outer product matrix multiplication.
 The newly developed algorithm called PB-SpGEMM saturates memory bandwidth by using the propagation blocking technique and by performing in-cache sorting and merging. 
 For many practical matrices, PB-SpGEMM runs 20\%-50\% faster than the state-of-the-art heap and hash SpGEMM algorithms on modern multicore processors.  
 Most importantly, PB-SpGEMM attains performance predicted by the Roofline model, and its performance remains stable with respect to matrix size and sparsity.  
 %Since existing algorithms do not attain even the lower bounds because of irregular memory accesses, 

\end{abstract}

\section{Introduction}
Sparse matrix-matrix multiplication (SpGEMM) is a widely-used kernel in many graph analytics, scientific computing, and machine learning algorithms. In graph analytics, SpGEMM is used in betweenness centrality \cite{hypersparse}, clustering coefficients, triangle counting \cite{triangles}, multi-source breadth-first search \cite{gapdt}, colored intersection search \cite{colored-intersection-searching}, and cycle detection \cite{cycle-detection} algorithms. In scientific computing, SpGEMM is used in algebraic multigrid \cite{ballard2016reducing} and linear solvers. Many machine learning tasks like dimensionality reduction (e.g., NMF, PCA \cite{fast-pca}), spectral clustering \cite{Jin2016AHP}, and Markov clustering (MCL) \cite{HipMCL}) rely on an efficient SpGEMM algorithm as well. Additionally, SpGEMM algorithms are applied to evaluate the chained product of sparse Jacobians \cite{Griewank-Jacobians} and optimize join operations on modern relational databases \cite{fast-join}.

In most data analytics applications, SpGEMM has very low arithmetic intensity (AI) measured by the ratio of total floating-point operations to total data movement.
For example, when multiplying two \erdosrenyi~random matrices with $d$ nonzeros per column (matrices with $d$ nonzeros uniformly distributed in each column) , an algorithm has an AI of just $\frac 1 {16}$ flops/byte (see Sec.~\ref{sec:ai}).
At this arithmetic intensity, SpGEMM is a memory-bound operation, and SpGEMM's peak performance has an upper bound of $\beta * AI$, where $\beta$ is the memory bandwidth. 
Assuming 50GB/s bandwidth available on a multicore processor, the estimated peak performance can be as high as $50/16 = 3.13$ GFLOPS (billions of floating point operations per second). 
However, state-of-the art parallel algorithms  based on heap and hash merging~\cite{nagasaka2019performance} attain no more than 500 MFLOPS on a socket of an Intel Skylate processor.  
No prior work clearly explained these observed performances of SpGEMM algorithms as no standard performance model has been developed to understand SpGEMM's performance. 

In this paper, we rely on the Roofline model~\cite{williams2009roofline} and develop lower and upper performance bounds for SpGEMM algorithms. 
In developing practical lower bounds of AI, we considered the fact that an SpGEMM may read data more than once.
Even with a tight lower bound on AI, an algorithm can attain peak performance (that is $\beta * AI$) only if it (a) saturates the memory bandwidth, (b) does not have high latency cost, and (c) makes use of full cache lines of data.
We show that these requirements are not satisfied by current column-by-column algorithms, resulting into lower than attainable peak FLOPS.  

Here, we develop a new algorithm based on the outer product of matrices.
The goal of this algorithm is to eliminate irregular data accesses, increase bandwidth utilization, and attain the performance predicted by the Roofline model. 
%Following the expansion-sort-compress We developed a new algorithm based on outer product multiplication.
Our algorithm is built upon the expansion-sort-compress paradigm~\cite{bell2012siam, dalton2015toms}.
Given two $n\times n$ matrices, this algorithm performs $n$ outer products to generate intermediate tuples of row index, column index, and multiplies values. 
These tuples are then sorted, and duplicate row and column indices are merged to get the final product. 
To ensure efficient bandwidth utilization, we store intermediate tuples into partially-sorted bins so that each bin can be sorted and merged independently.
This technique of binning intermediate data for better bandwidth utilization is called \emph{propagation blocking} (PB)~\cite{beamer2017reducing}.
Prior work has used propagation blocking to improve the bandwidth utilization as well as the overall performance of sparse matrix-vector multiplications~\cite{beamer2017reducing, ozdal2019improving}.
Here we use propagation blocking to regularize data movements in SpGEMM. 
Hence, our algorithm is named as \emph{PB-SpGEMM}.

All three phases of PB-SpGEMM (expand, sort, and compress or merge) stream data from memory.
Hence, every phase has no significant latency overhead, utilizes full cache lines, and attains a bandwidth close the STREAM benchmark.
Therefore, given the bandwidth of a system and input matrices, PB-SpGEMM's performance matches the prediction of the Roofline model.

Similar to any expansion-sort-compress algorithm, PB-SpGEMM has to store $\flop$ intermediate tuples, where  $\flop$ is the number of multiplications needed by SpGEMM.
This can lead to significant data movements when the \emph{compression factor}  (the ratio of $\flop$ to nonzeros in the output) is greater than four.
However, most practical SpGEMM operations have small compression factors.
For example, when squaring matrices from the SuiteSparse Matrix Collection,  more than $80\%$ of SpGEMMs have a compression factor less than 3 and about $99\%$ have a compression factor less than 6~\cite{liu2019register}.
Hence, for most practical scenarios, PB-SpGEMM performs predictably better than existing heap and hash algorithms. 
For multiplications with compression factor greater than four, PB-SpGEMM's performance is still predictable, but it can run slower than the alternatives.

We summarize the key contributions of this paper below:
\begin{enumerate}
    \item We develop a Roofline performance model for SpGEMM algorithm. This model can show the limitations of existing column-by-column SpGEMM algorithms. 
    \item We develop an outer-product based SpGEMM algorithm called PB-SpGEMM. We use propagation blocking, in-cache sorting and merging for better bandwidth utilization. 
    \item Different phases of PB-SpGEMM utilize bandwidth close to the STREAM benchmark. Given the bandwidth of a system and input matrices, PB-SpGEMM's performance matches the prediction from our Roofline model.
    \item For SpGEMMs with compression factor less than four,  PB-SpGEMM is 30\% to 50\% faster than previous state-of-the-art algorithms for multicore processors. 
\end{enumerate}

Our implementation of PB-SpGEMM is publicly available at Bitbucket: https://bitbucket.org/azadcse/outerspgemm .
\section{A Performance Model for SpGEMM}
\subsection{Notations}
Given two sparse matrices $\mA {\in} \mathbb{R}^{m \times k}$ and ${\mB {\in} \mathbb{R}^{k \times n}}$, SpGEMM computes another potentially sparse matrix ${\mC {\in} \mathbb{R}^{m \times n}}$.
$\mA(:,j)$ denotes $j$th column of $\mA$, and $\mA(i,j)$ denotes $i$th entry in the $j$th column. 
In our analysis, we consider $n$-by-$n$ matrices for simplicity.
We use \erdosrenyi~(ER) random matrices throughout the paper.
An ER matrix with $d$ nonzeros per column has $d$ nonzeros uniformly distributed in each column.
Among many representations for sparse matrices, we consider three standard data structures in this paper: Compressed Sparse Row (CSR), Compressed Sparse Column (CSC), and Coordinate format (COO). See Langr and Tvrdik ~\cite{langr2015evaluation} for a comprehensive discussion on sparse matrix storage formats.

Given a matrix $\mA$, $\nnz(\mA)$ denotes the number of nonzeros in $\mA$, and ${d(A)=\frac{\nnz(\mA)}{n}}$ denotes average nonzeros in a row or column.
In computing $\mC{=}\mA\mB$, $\flop$ denotes the number of multiplications needed.
Throughout the paper, floating point operations only denote multiplications.
The \emph{compression factor} ($\cf$) denotes the ratio of $\flop$ to nonzeros in the output matrix: $\cf {=} \frac{\flop}{\nnz(\mC)}$. 
Since at least one multiplication is needed for every output nonzero, $\cf{\geq} 1$.

\subsection{Classes of SpGEMM algorithms categorized by data access patterns}
To compute $\mC{=}\mA\mB$, most algorithms also manipulate an expanded matrix $\hat{\mC}{\in} \mathbb{R}^{m \times n}$ that contains $\flop$ unmerged entries.
The final output $\mC$ can be obtained by merging entries in $\hat{\mC}$ with the same row and column indices. 
Therefore, if we primarily focus on data accesses, SpGEMM algorithms have two distinct phases: (a) input matrices are accessed to form $\hat{\mC}$, (b) duplicate entries in $\hat{\mC}$ are merged to form $\mC$.
Here, merging means adding multiplied values with the same row and column indices. 
For input accesses, we have only two options: (1) access $\mA$ and $\mB$ column-by-column (or equivalently row-by-row)~\cite{nagasaka2019performance, patwary2015parallel, deveci2017performance, azad2016exploiting, dalton2015toms}, and (2) access $\mA$ column-by-column and access $\mB$ row-by-row for outer product~\cite{buluc2008representation, OuterSPACE-8327050}.
For output formations, we have many options. Prior work used the expand-sort-compress strategy~\cite{dalton2015toms, OuterSPACE-8327050, liu2019register} or used accumulators based on heap~\cite{azad2016exploiting}, hash table~\cite{nagasaka2019performance}, and a dense vector called SPA~\cite{patwary2015parallel, gilbert1992sparse}. 
Table~\ref{tab:algo-classifiication} summarizes prior work based on their data access patterns.
Next, we briefly summarize four major classes of algorithms and discuss their data access patterns.

\begin{table}[!t]
    \centering
    \begin{adjustbox}{max width=\linewidth}
    \begin{tabular}{@{} l | l c c  @{} }
    \toprule
    & & \multicolumn{2}{c}{Input Access} \\
    \toprule
    & & Column wise & Outer Product\\
    Output  & Heap/Hash/SPA & \cite{nagasaka2019performance, patwary2015parallel, deveci2017performance, azad2016exploiting} & \cite{buluc2008representation}\\
    Formation & ESC & \cite{dalton2015toms, liu2019register} & \cite{OuterSPACE-8327050}, this\\
    \bottomrule

    \end{tabular}
    \end{adjustbox}

    \caption{Classification of SpGEMM algorithms based on access patterns of input and output matrices. This paper falls into the lower right cell.}

    \label{tab:algo-classifiication}
\end{table}

\begin{figure}[!t]
    \centering
    \includegraphics[width=0.9\linewidth]{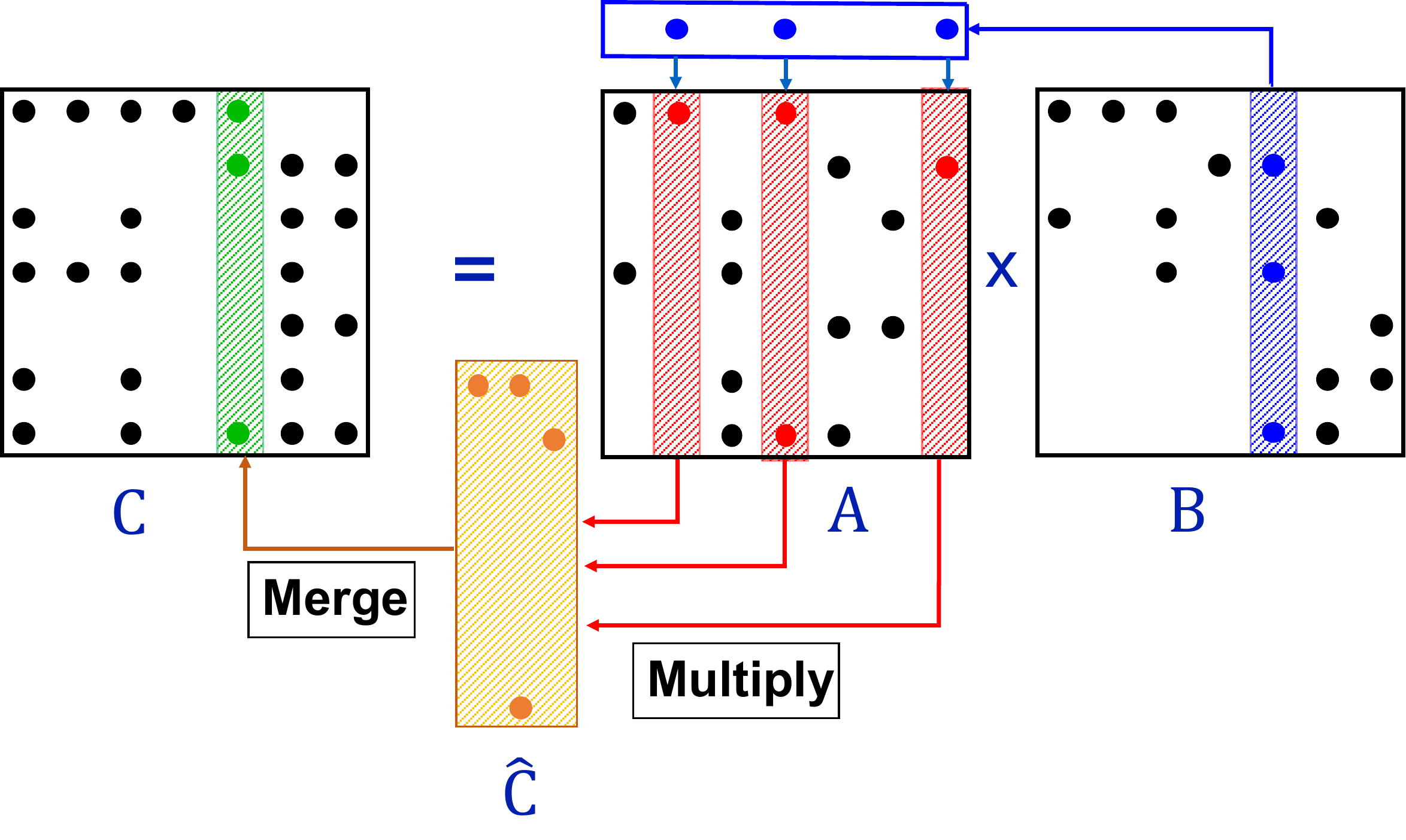}
    \caption{An illustration of column SpGEMM algorithm. To generate $\mC(:,5)$, the algorithm selects 2nd, 4th and 7th columns of $\mA$ (shown in red) based on the nonzeros in $\mB(:,5)$ (shown in blue).  The selected columns are multiplied by corresponding entries in $\mB(:,5)$, forming entries in $\hat{\mC}(:,5)$ (shown in yellow). Duplicate entries in $\hat{\mC}(:,5)$ are merged to generate $\mC(:,5)$ (shown in green). The merging strategy differs in different column SpGEMM algorithm}
    \label{fig:ColumnSpGEMM}
\end{figure}

\begin{figure*}[t]
    \centering
    \includegraphics[width=0.8\linewidth]{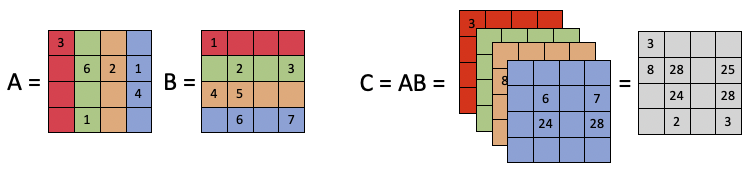}
 
    \caption{An illustration of outer product SpGEMM algorithm. Each multiplication between a column from A and a corresponding row from B generate a sub 1-rank matrix, those sub 1-rank matrices added up yield the final result C}
    \label{fig:illustration-outerproduct}
\end{figure*}

\begin{table*}[!t]
    
    \centering
    \begin{tabular}{@{} l l l l l | l l l l | c c c c@{} }
    \toprule
    & \multicolumn{4}{c|}{No of Accesses} & \multicolumn{4}{c|}{Streamed Access} & \multicolumn{4}{c}{Cache Line Utilization}\\
    %\cmidrule(l){2-5} \cmidrule(l){6-9} \cmidrule(l){10-13}
    \midrule
    Algorithm & $\mA$ & $\mB$ & $\hat{\mC}$ & $\mC$ &   $\mA$ & $\mB$ & $\hat{\mC}$ & $\mC$  &   $\mA$ & $\mB$ & $\hat{\mC}$ & $\mC$ \\
     \toprule
    Column SpGEMM (Heap/Hash/SPA)& $d$ & $1$ & $0^{*}$ & $1$ & $\times$  & \checkmark & \checkmark & \checkmark & $\times$ (when $d<8$ ) & \checkmark & \checkmark & \checkmark \\
    ESC (column-wise)& $d$ & $1$ & $2$ & $1$ & $\times$  & \checkmark & $\times$ & \checkmark  & $\times$ (when $d<8$ ) & \checkmark &  \checkmark & \checkmark\\
    ESC (outer product) & $1$ & $1$ & $2$ & $1$ & \checkmark  & \checkmark & $\checkmark^{**}$ & \checkmark & $\checkmark$  & \checkmark & \checkmark & \checkmark \\
    
    \bottomrule

    \end{tabular}
     \\
    \textsuperscript{*} {\footnotesize Column SpGEMM generates one column of $\hat{\mC}$ at a time.}
    \textsuperscript{**} {\footnotesize Using blocking techniques discussed in this paper.}
  
    \caption{Data access patterns in different SpGEMM algorithms when multiplying two ER matrices with $d$ nonzeros per column. If an algorithm does not stream data, it has high latency cost. }

    \label{tab:data_access}
\end{table*}

{\bf Column SpGEMM algorithms based on the heap/hash/SPA accumulator.} 
Some algorithms materialize only one column of $\hat{\mC}$, merge duplicate entries in that column and generate the corresponding column of $\mC$.  
Since column-by-column and row-by-row algorithms have similar computational patterns, we only discuss column SpGEMM algorithms in this paper. 
An illustration of the column SpGEMM algorithm is shown in Fig.~\ref{fig:ColumnSpGEMM}, where $\mC(:,5)$ is generated by merging a subset of columns in $\mA$ determined by nonzeros in $\mB(:,5)$.
Most column-by-column algorithms are based on Gustavson's algorithm~\cite{gustavson1978two}, and they differ from one another based on how they merge entries in $\hat{\mC}(:,j)$ to obtain $\mC(:,j)$.
Prior work have used heap~\cite{azad2016exploiting}, hash table~\cite{nagasaka2017high}, or a dense vector called SPA~\cite{gilbert1992sparse} for merging columns.
A common characteristic of all column-by-column algorithms is that they read from $\mB$ and write to $\mC$ one column at a time. 
However, columns of $\mA$ may be read irregularly several times based on the nonzero pattern of $\mB$.

We explain the access pattern of $\mA$ when multiplying two ER matrices with $d$ nonzeros per column.
In the worst case, each column of $\mA$ will be read $d$ times from memory because columns of $\mA$ are accessed randomly without any spatial locality. Thus, we read $\mA$ $d$ times over the execution of the algorithm. 
Second, if we store matrices in the CSC format, column SpGEMM algorithms have good spatial locality for $\mB$, $\hat{\mC}$ and $\mC$, but not for $\mA$. Hence, we pay heavy latency costs for irregular accesses of $\mA$'s columns.
Third, if a column of $\mA$ has very few entries (e.g., when $d$ is less than 8), we read a column of $\mA$ in a cache line, but the whole cache line is not used. Hence, column SpGEMM waste memory bandwidth for very sparse matrtices.  
The first row of Table~\ref{tab:data_access} summarizes the data access patterns in column SpGEMM algorithms. 
Similarly, a row-by-row algorithm (when matrices are stored in the CSR format) has good spatial locality for $\mA$, $\hat{\mC}$ and $\mC$, but not for $\mB$.

{\bf The Expand-Sort-Compress (ESC) algorithms.} 
Algorithms based on the ESC technique generate full $\hat{\mC}$ before merging duplicate entries. 
The original ESC algorithm~\cite{dalton2015toms} developed for GPUs generates $\hat{\mC}(:,i)$ using $\mB(:,i)$\footnote{Earlier work~\cite{dalton2015toms} actually used a row-by-row algorithm with CSR matrices, which is equivalent to column SpGEMM with CSC matrices.} 
similar to the first step of column SpGEMM algorithm shown in Fig.~\ref{fig:ColumnSpGEMM}.
After the entire $\hat{\mC}$ is constructed, $\flop$ tuples in $\hat{\mC}$ are sorted and merged to generate the final output. 
Since sorting can be efficiently performed on GPUs, ESC SpGEMM can performed better than other algorithms on GPUs~\cite{dalton2015toms, liu2019register}.
%(recent work suggest that hash-based column SpGEMM performs the best on GPUs~\cite{}). 
The column ESC algorithm has access patterns for $\mA$, $\mB$, and $\mC$ similar to the column SpGEMM algorithm. 
Additionally,  $\hat{\mC}$ needs to access twice (one write after multiplication and one read before merging).
The second row of Table~\ref{tab:data_access} summarizes the data access patterns in the column ESC algorithms. 

Previous work~\cite{dalton2015toms} tried to eliminate reading and writing $\hat{\mC}$ from memory by partitioning $\mA$ by rows and multiplying each partition with $\mB$.
Thus, this approach generates one partition of $\hat{\mC}$ at a time, which can fit in cache when a large number of partitions is used.
However, the partitioned ESC algorithm will need to read  $\hat{\mB}$ several times as well as reading $\hat{\mA}$ multiple times for each partition. 
Hence, the effectiveness of partitioning depends on the number of partitions and nonzero structures of input matrices.

{\bf Outer product algorithms.} 
Fig.~\ref{fig:illustration-outerproduct} shows an illustration of the outer product algorithm.
In this formulation, $A(:,i)$ is multiplied with $B(i,:)$ to form a rank-1 outer product matrix. 
The rank-1 matrices can be merged by using a heap or using ESC.
A heap can be used to merge the outer product of $A(:,i)$ and $B(i,:)$ with the current output after every iteration~\cite{buluc2008representation}. However, this algorithm is too expensive as it requires $n$ merging operations. Hence, we do not elaborate this algorithm further.    

The rank-1 matrices from outer product can also be merged using the ESC strategy. 
In this case, input matrices are streamed only once. 
Hence, outer product can fully utilize cache lines when reading inputs.  
To sort and merge unmerged tuples in $\hat{\mC}$, we need to read them from memory. 
This can significantly increase the memory traffic. 
Nonetheless, with an efficient blocking technique discussed in this paper, we can stream $\hat{\mC}$ whenever it is accessed. 
Hence, we can utilize full cache lines  and saturate memory bandwidth to offset more data accesses. 
The last row of Table~\ref{tab:data_access} summarizes the data access patterns in the outer-product-based ESC algorithm.

\subsection{Arithmetic Intensity (AI) of SpGEMM.} 
\label{sec:ai}
AI is the ratio of total floating-point operations to total data movement (bytes).
To compute $\mC{=}\mA\mB$, one must read $\mA$ and $\mB$ from memory, and write $\mC$ to memory\footnote{We ignore that $\mC$ may have to first read from memory to cache before writing.}.
Assume that on average, we need $b$ bytes to store a nonzero.
Then,
\begin{equation}
    \text{AI} \leq \frac{\nnz(C) * \cf}{[\nnz(A) + \nnz(B) + \nnz(C)]*b } \leq \frac{\cf}{b}.
    \label{eq:AI_upper}
\end{equation}
If we use 4 bytes for indices and 8 bytes for values, then $b$ is 16 bytes (assuming that matrices are stored in the COO format). 
Here, $\cf$ is a property of input matrices and it varies from 1 to 8 for most practical sparse matrices~\cite{liu2019register}.
%Given these assumptions on $b$ and $\cf$, Equation\ref{eq:AI} reveals several important aspects of SpGEMM.
Even in the best scenario when we read and write matrices just once, the arithmetic intensity of SpGEMM is very low: often less than one.
Let's consider multiplying two ER matrices $A$ and $B$, since $\cf$ for ER matrix is close to 1 in expectation~\cite{ballard2013communication}, according to our model, AI will be around $\frac 1 {16}$ flops/byte.
At this AI, SpGEMM's performance is completely bound by memory bandwidth\footnote{On modern processors, SpGEMM can be computation bound only if $\cf>1000$, it is unrealistic for sparse matrices.}.

\begin{figure}[!t]
    \centering
    \includegraphics[width=0.9\linewidth]{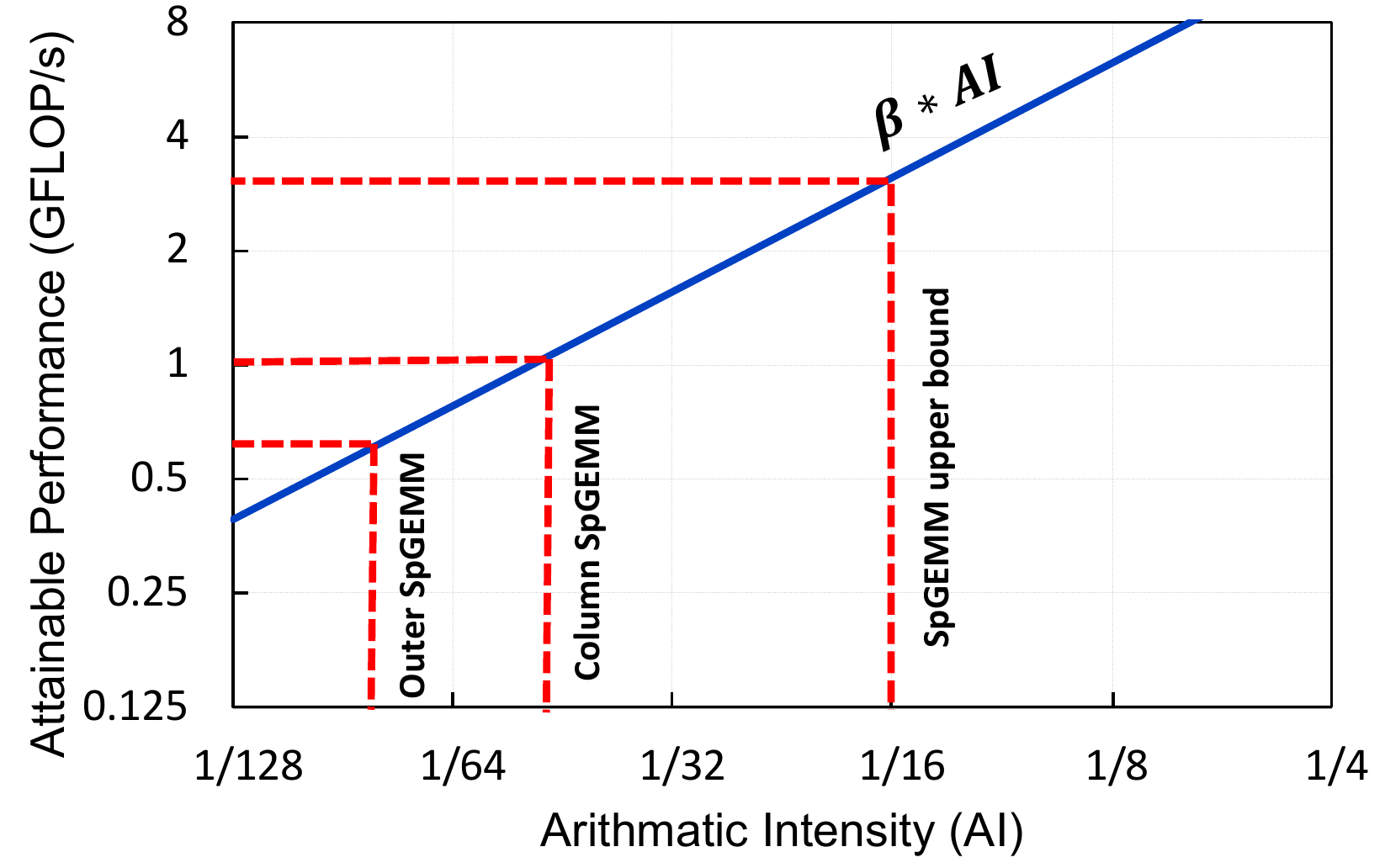}
   
    \caption{Using Roofline model to estimate performance when multiplying two \erdosrenyi ~matrices on a single-socket Intel Skylake processor. Memory bandwidth $\beta$ is 50GB/s as measured with STREAM benchmark}
 
    \label{fig:roofline}
\end{figure}

{\bf Peak  performance of an SpGEMM algorithm.}
Suppose, a smart algorithm achieves the best AI of $\frac{\cf}{b}$.
Let $\beta$ be the memory bandwidth (BW) of the system measured by the STREAM benchmark.
Then, the performance measured by FLOPS (floating point operations per second) follows this inequality:
\begin{equation}
\text{FLOPS} \leq \beta \frac{\cf}{b}.
\end{equation}
Hence, the \emph{peak} FLOPS for a given problem on a given architecture can be at most $\beta \frac{\cf}{b}$ assuming that SpGEMM is bandwidth bound. 
For example, on an Intel Skylake processor with 50GB/s memory bandwidth, the peak performance for multiplying ER matrices can be at most $50*1/16=3.13$ GFLOPS as shown in Fig.~\ref{fig:roofline}.  
However, state-of-the-art column SpGEMM algorithms achieve less than 20\% of this peak performance as discussed in several recent papers~\cite{nagasaka2019performance}. 

As discussed before, the primary reasons behind the suboptimal performance of SpGEMM algorithms are: (1) algorithms read/write data multiple times, (2) algorithms access data at random memory locations, (3) algorithms may not fully utilize cache lines. 
The first problem is inherent to SpGEMM and cannot be completely overcome when input matrices are unstructured.
The irregular data access can under utilize bandwidth, impeding the performance of SpGEMM when the compression factor is small. 
In this paper, we address this irregular access problem and develop an algorithm where all steps of SpGEMM utilize full memory bandwidth. 
Nevertheless, Equation~\ref{eq:AI_upper} seems an upper bound that no existing algorithm can attain. 
Next, we consider a more practical bound for AI.

{\bf A more practical bound on AI for SpGEMM.}
In column SpGEMM algorithm, we read the first input matrix several times depending on the nonzero pattern of $\mB$. 
To obtain a lower bound for column SpGEMM, we assume that the accesses of $\mA$ have no temporal or spatial locality, and all accesses of $\mA$ incur memory traffic.
Hence, in the worst case, the amount of data read from $A$ is $\flop * b$ bytes. Hence AI for column SpGEMM can be approximated as follows:
\begin{eqnarray}
\text{AI(col)} & \geq & \frac{\nnz(C) * \cf}{[\nnz(C) * \cf + \nnz(B) + \nnz(C)]*b } \nonumber\\
             & \geq & \frac{\cf}{(2+cf)b}.
\label{AI_column}
\end{eqnarray}
By contrast, the outer product based algorithm based on the ESC strategy generates all unmerged tuples in $\hat{\mC}$, writes all \nnz($\hat{\mC}$) = $\flop$ tuples in memory, reads them again for sorting and merging.
Hence, in the worst case, ESC-based algorithms performs additional $2*\flop$ memory red-write operations, giving us the following AI:
\begin{eqnarray}
\text{AI(outer)} & \geq & \frac{\nnz(C) * \cf}{[\nnz(\mA) + \nnz(\mB) +\nnz( \hat{\mC} ) + \nnz(\mC)]*b } \nonumber\\
& = & \frac{\nnz(C) * \cf}{[\nnz(A) + \nnz(B) + 2*\flop + \nnz(C)]*b } \nonumber\\
             & \geq & \frac{\cf}{(3+2\cf)b}.
\label{AI_outer}
\end{eqnarray}
For ER matrices with $\cf=1$, Eq.~\ref{AI_outer} gives us an arithmetic intensity of 1/80 (assuming $b=16$ bytes).
Fig.~\ref{fig:roofline} shows these lower bounds on AI with the corresponding attainable performance. 
We will experimentally show that the newly-developed outer-product-based algorithm can attain the peak performance based on Eq.~\ref{AI_outer}.

% \subsection{Effective memory bandwidth of SpGEMM algorithms.}
% Since different SpGEMM algorithms access data differently (both in number of passes and regularity in access), it is hard to compare sustained bandwidths attained by different algorithms.  
% We instead consider a measure of effective bandwidth $\Tilde{\beta}$

% \begin{equation}
%     \Tilde{\beta} = \frac{[\nnz(A) + \nnz(B) + 2*\nnz(C)]*b } {t_s}, 
% \end{equation}
% where $t_s$ is the runtime of an algorithm.
% $\Tilde{\beta}$ is fair to algorithms since the numerator is problem and data structure dependent. 
% The runtime $t_s$ is inversely proportional to  $\Tilde{\beta}$, which is expected in bandwidth-bound algorithms. 
% If an algorithm read data several times or access memory irregularly, its effective bandwidth $\Tilde{\beta}$ will be much lower than STREAM bandwidth $\beta$.
% This performance measurement is motivated by Graph500 benchmark where traversed edge per seconds (TEPS) is measured by dividing the number of edges in a graph by the runtime of an algorithm. 
% In Graph500, an algorithm can achieve better TEPS by traversing fewer edges (e.g., using the direction-optimized search~\cite{}).
% Here, a SpGEMM algorithm can achieve better $\Tilde{\beta}$ by accessing data contiguously and utilizing memory hierarchy efficiently.

%\subsection{Irregular data accesses in current SpGEMM algorithms and its impact on effective memory bandwidth}

\section{The PB-SpGEMM Algorithm}

\begin{figure*}[!t]
    \centering
    \includegraphics[width=0.8\linewidth]{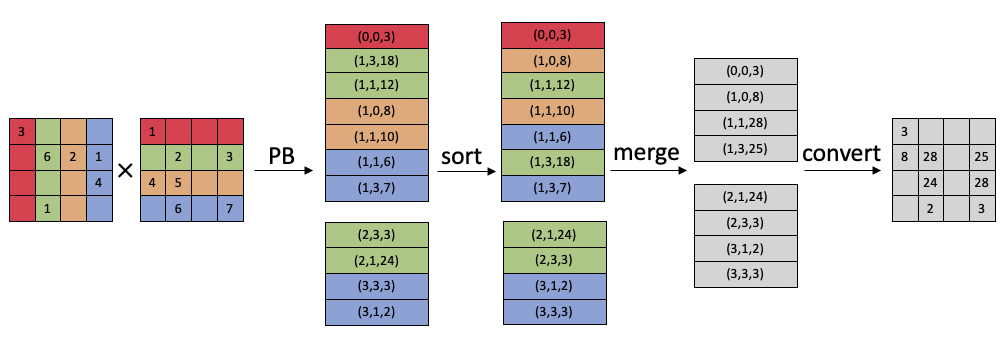}
    %\vspace{-10pt}
    \caption{An example of PB-SpGEMM multiplying two 4$\times$4 matrices with two bins, the first bin blocks propagation of tuples with rowid 0 and 1, the second blocks 2 and 3}
    %\vspace{-10pt}
    \label{fig:illustration-PB}
\end{figure*}

\begin{figure}[!t]
    \centering
    \includegraphics[width=0.9\linewidth]{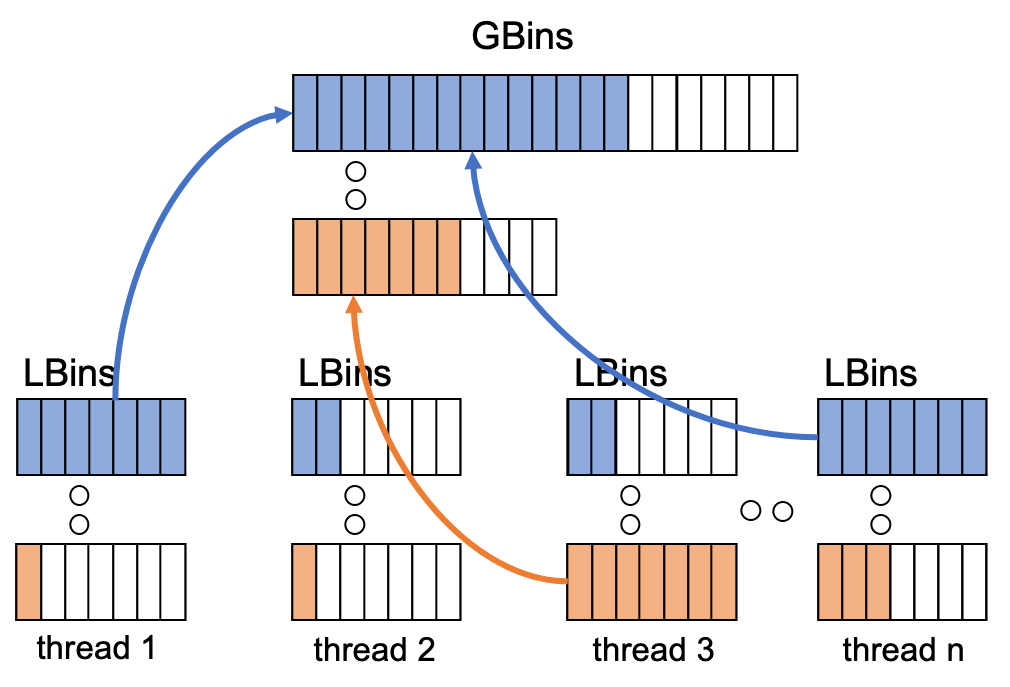}
    %\vspace{-10pt}
    \caption{Using thread-private local bins to utilize cache lines. If $nbins$ is the total number of global bins, each thread also creates $nbins$ number of small local bins to store tuples as they are generated. When a local bin is full, tuples inside will be flushed to the corresponding global bin. This example is using two local bins per thread and two global bins}
    %\vspace{-5pt}
    \label{fig:localbins}
\end{figure}

\subsection{Overview of the PB-SpGEMM algorithm}
Algorithm~\ref{algo:esc-spgemm} provides a high-level description of an SpGEMM algorithm based on the expand-sort-compress scheme. 
In the symbolic step, we estimate $\flop$ for the current multiplication and allocate memory for unmerged tuples in $\hat{\mC}$.
Then, multiplied tuples are formed and stored in  $\hat{\mC}$, which is then sorted and merged to form $C$.

Our algorithm follows the exact same principle, but uses outer products and propagation blocking for efficient bandwidth utilization. 
Fig.~\ref{fig:illustration-PB} explains the propagation blocking idea based on two $4\times 4$ matrices and two bins. 
After we expand tuples, we partially order them in two bins, where the first bin stores rowid 0 and 1, and the second bin stores rowid 2 and 3. 
If these bins fit in L2 cache, sorting and merging can be be performed efficiently in cache by different threads. 
After generating a tuple, if we directly write it to its designated global bin, we may not fully utilize the cache line. 
Hence, each thread also maintains small local bins that are filled in cache before flushing to global bins in memory.
The use of local bins is illustrated in Fig.~\ref{fig:localbins}.
Hence, our algorithm has several tunable parameters, including (a) nbins: number of global or local bins, and (b) Lbinwidth: the width of local bins. We experimentally select these parameters as will be discussed in the experimental section.

Algorithm~\ref{algo:pb-spgemm} described the PB-SpGEMM algorithm. 
To facilitate outer product operations, input matrices $\mA$ and $\mB$ are passed as CSC and CSR formats, respectively.
Here, we store $\mC$ in CSR format, but it can be easily stored in CSC without any overhead. 
Finally, the expanded matrix $\hat{\mathbf{C}}$ is stored in the COO format. 
Similar to Algorithm~\ref{algo:esc-spgemm}, PB-SpGEMM has four phases (a) symbolic (line 1) (b) Expand (lines 5-14), (c) Sort (line 16), and (d) Compress (line 17). 
However, Algorithm~\ref{algo:pb-spgemm} differs from previous ESC algorithms in two crucial ways: (1) we use outer product to stream data from input matrices, (2) we use propagation blocking to organize the expanded matrix into bins so that all phases of the algorithm saturate the memory bandwidth.
We additionally perform a post-processing step (line 9) to convert the output to CSR format. 
Next, we discuss these phases in detail. 

\begin{algorithm}[!t]

%\caption{OuterSpGEMM algorithm}
\LinesNumbered
\KwIn{$\mathbf{A}$ , $\mathbf{B}$}
\KwOut{$\mathbf{C}$ }
\SetKwFunction{Symbolic}{Symbolic}
\SetKwFunction{Expand}{Expand}
\SetKwFunction{Append}{PBAppend}
\SetKwFunction{PB}{PB}
\SetKwFunction{Sort}{Sort}
\SetKwFunction{Compress}{Compress}
\SetKwFunction{Convert}{ConvertCSR}

$\hat{\mathbf{C}} \gets $\Symbolic(A, B)\Comment{ Create space for $\hat{\mathbf{C}}$}\;
     $\hat{\mathbf{C}} \gets$  \Expand($\mathbf{A}$, $\mathbf{B}$) \Comment{Create unmerged tuples} \;
 \Sort($\hat{\mathbf{C}}$) \Comment{perform radix sort using (rowid, colid) as keys}\;
    $\mathbf{C}$ $\leftarrow$ \Compress($\hat{\mathbf{C}}$) \Comment{merge duplicated tuples }\;

 \caption{ESC-SpGEMM algorithm}

 \label{algo:esc-spgemm}
\end{algorithm}
 %\vspace{-10pt}
 
\begin{algorithm}[!t]

%\caption{OuterSpGEMM algorithm}
\LinesNumbered
\KwIn{$\mathbf{A}$ in CSC, $\mathbf{B}$ in CSR}
\KwOut{$\mathbf{C}$ in CSR}
\SetKwFunction{Symbolic}{Symbolic}
\SetKwFunction{Expand}{Expand}
\SetKwFunction{Append}{Append}
\SetKwFunction{MemCopy}{MemCopy}

\SetKwFunction{PB}{PB}
\SetKwFunction{Sort}{Sort}
\SetKwFunction{Compress}{Compress}
\SetKwFunction{Convert}{ConvertCSR}

GBin $\leftarrow \Symbolic(\mA, \mB)$ \Comment{ expanded tuples ($\hat{\mC}$) partitioned into $nbins$ bins}\;

nthreads $\gets$ Number of threads \;
Lbinwidth $\gets$ 512  \Comment{Local bin width, default set to 512 bytes} \;
LBin $\leftarrow$ Create a 3D array of size nthreads $\times$ nbins $\times$ Lbinwidth \Comment{LBin[$T_i$] is local bin, private for thread $T_i$} \;

%$\hat{\mathbf{C}}$ $\leftarrow \phi$ \Comment{ expanded tuples partitioned into $nbins$ bins}\;

 \For{$i \leftarrow 1$ \KwTo $k$ in parallel} 
 {
    %$\hat{\mathbf{C}_1}$ $\leftarrow$ \Expand($\mathbf{A}(:, i)$, $\mathbf{B}(i, :)$) \Comment{generate multiplied tuples}\;
    %$\hat{\mathbf{C}} \leftarrow \Append(\hat{\mathbf{C}}, \hat{\mathbf{C}_1})$ \Comment{append to $\hat{\mathbf{C}}$ by  concatenating tuples to global bins}\;
    $T_i \leftarrow$ The thread executing the $i$th iteration \;
    %\For{$i \leftarrow 1$ \KwTo $k$ in parallel} 
    \For{(rowid, i, aval) $\in \mA(:, i)$}
    {
        \For{(i, colid, bval) $\in \mB(i:,)$}
        {
            binid $\gets$ rowid \% nbins \;
            
            \If{size(LBin[$T_i$][binid]) = Lbinwidth}{
              \MemCopy(GBin[binid], LBin[$T_i$][binid]) \;
              LBin[$T_i$][binid] $\gets \phi$ \;
            }
            mval $\gets$ aval $\times$ bval \;
            \Append(LBin[$T_i$][binid], (rowid, colid, mval))\;
        }
    }
}

\For{all thread $T_i$  in parallel} 
{
\For{binid $\gets$ 1 \KwTo $nbins$} 
{
    \If{size(LBin[$T_i$][binid]) $\neq$ 0}
    {
              \MemCopy(GBin[binid], LBin[$T_i$][binid]) \;
    }
}
}
\For{binid $\leftarrow 1$ \KwTo $nbins$ in parallel} {
    GBin[binid] $\leftarrow$ \Sort(GBin[binid]) \Comment{perform radix sort in the $i$th bin using (rowid, colid) as keys}\;
    GBin[binid] $\leftarrow$ \Compress(GBin[binid])
    \Comment{merge duplicated tuples by two-pointer method}\;
 }
 $\mathbf{C} \leftarrow \Convert(GBin)$\;
 \caption{PB-SpGEMM algorithm}
 \label{algo:pb-spgemm}
\end{algorithm}

\subsection{Symbolic Phase}
In the symbolic phase, we estimate the memory requirement for $\hat{\mathbf{C}}$, estimate number of bins and allocate space for global bins (Gbin).
Algorithm~\ref{algo:symbolic} describes the symbolic step. 
We compute flops for the current multiplication using an outer-product style computation.
The loop in Line 2 accesses $\mA$ column by column and $\mB$ row by row.
If there is a nonzero entry in $\mA(rowid,i)$, it must be multiplied by all nonzeros in the $i$th row of $\mB$.
Hence, line 5 adds $\nnz (\mB(i,:)) \times \nnz (\mA(: , i))$ to the $\flop$ count.
After we compute flops, we compute the number bins (line 6) so that each global bin fits in the L2 cache in the sorting and merging phase.
We then allocate memory for global bins (line 7).  Algorithm~\ref{algo:symbolic} needs only $O(n)$ time and attains higher memory bandwidth by streaming just row and column pointer arrays of $\mB$ and $\mA$, respectively.

% If there is a nonzero entry in $\mA(rowid,i)$, it must be multiplied by all nonzeros in the $i$th row of $\mB$.
% Hence, line 7 adds $\nnz (\mB(i,:))$ to the $\flop$ count of the corresponding row.
% Algorithm~\ref{algo:symbolic} needs $O(\nnz(\mA))$ time to populate the $flops$ array. 
% It only touches all row indices of $A$, but does not need column indices of $B$. 
% Hence, Algorithm~\ref{algo:symbolic} can attain higher memory bandwidth by streaming $\mA$ stored in the CSC format.

Note that Algorithm~\ref{algo:symbolic} is much simpler than symbolic steps used in column SpGEMM algorithms where we need to estimate $\nnz(\mC)$. 
An outer product algorithm can be also developed without a symbolic phase.
For example, a linked-list can be used to dynamically append expanded tuples~\cite{OuterSPACE-8327050}. However, the dynamic memory allocations in parallel sections could result in poor performance.

\begin{algorithm}[!t]
\LinesNumbered
\caption{Symbolic Phase (Static Schedule)}

\label{algo:symbolic}
\KwIn{A in CSC, B in CSR}
\KwOut{Gbin, nbins}
\SetKwFunction{MemAlloc}{MemAlloc}

% \For{$i$ $\leftarrow$ $1$ \KwTo $m$ in parallel} {
%     $flops[i]$ $\leftarrow$ 0\;
% }

flops $\gets$ 0\;

\For{$i$ $\leftarrow$ $1$ \KwTo $k$ in parallel} {
  
  $\nnz (\mB(i,:))$ $\leftarrow$ $\mB.rowptr[i+1]$ - $\mB.rowptr[i]$\;
  $\nnz (\mA(:,i))$ $\leftarrow$ $\mA.colptr[i+1]$ - $\mA.colptr[i]$\;
  flops $+=$ $\nnz (\mB(i,:))  \times \nnz (\mA(:,i))$
%   \For{$j$ $\leftarrow$ $\mA.colptr[i]$ \KwTo $\mA.colptr[i+1]$} {
%     $rowid$ $\leftarrow$ $\mA.rowids[j]$\;
%     $flops[rowid]$ += $\nnz (\mB(i,:))$;
%   }
}

nbins $\gets$  flops / \texttt{L2\_CACHE\_SIZE} \; 
Gbin $\gets$ MemAlloc(flops) \Comment{allocate shared array to store tuples, no initialization needed}

\end{algorithm}

\subsection{Expand}
Lines 5-14 in Algorithm~\ref{algo:pb-spgemm} describes the expand phase of PB-SpGEMM. 
In the expand phase, a thread reads $\mA(;,i)$ and $\mB(i,:)$ and performs their outer product.
The binid is computed by rowid of the multiplied tuple (rowid comes from the rowid in $\mA(:,i)$).
Once a local bin, this thread will flush the tuples inside to the corresponding global bin (line 10-12).
Then, the newly-formed tuple is appended to its designated local bin in line 14.
After the multiplication, there still could be some tuples left in the local bins because bins were not full.
Lines 15-18 send the partially full local bins to global bins.

With the local binning, we always write tuples in multiple of cache lines.  We make sure the number of local bins and the size of local bins are small.
typically, we create 1024 bins and 512 bytes per local bin so that all local bins for a thread easily fit in the cache.

\subsection{Sorting} 
After the multiplication and propagation blocking phase, the expanded matrix $\hat{\mC}$ is stored in the COO format, partitioned into several bins.  
Then, we sort $\hat{\mC}$ to bring tuples with the same (rowid, colid) pair close to one another for merging in the next phase. 
As shown in Algorithm~\ref{algo:pb-spgemm}, sorting can be performed independently in each bin because bins do not share tuples with the same rowid.
Hence, a thread can sort tuples in a bin sequentially, while other threads sort other bins in parallel.

The sorting algorithm uses the (rowid, colid) pairs as keys and the multiplied values as payloads. 
For this purpose, we use an in-place radix sort (similar to American flag sort~\cite{mcilroy1993engineering}) that groups the keys by individual bytes sharing the same significant byte position. 
In the worst case, this in-place radix sort needs $b$ passes over the data, where $b$ is the number of bytes needed to store a key.
%The worst case complexity of Radix sort is $O(nw)$, where $w$ is the number of bits required to store each key, and $n$ is the number of keys. 
Hence, Radix sort can be faster than comparison-based sorting when keys are stored in fewer bytes. 

{\bf Preparing integer keys for Radix sort.}
In our algorithm, we use 4-byte integers for row and column indices. 
We concatenate the rowid and colid to form a combined 8-byte integer key for radix sort.
With 8-byte keys, radix sort may need 8 passes over the data to sort tuples, which can incur significant data transfers. 
We can reduce the key space by using the fact that bins are already grouped by consecutive row indices. 
For example, if the input is a 1M $\times$ 1M matrix and we create 1K bins to block the propagation, the rowids of tuples within the same bin will in adjacent 1k, then we only need 10 bits to represent the remainder rather than a full 32-bit integer, and we still have 32-10=22 bits to store colid.
Furthermore, if we assume that matrices have at most 1M rows and columns, we can use 20 bits for colid and 20-10=10 bits for rowid (assuming 1K bins).
Hence, in most practical cases, we can potentially squeeze keys into 4-byte integers, needing four passes over the data for sorting.

\begin{table*}[!t]
    
    \centering
    \begin{tabular}{@{} p{1cm} l l l l  l l @{} }
    \toprule
    Phase & Comp. Complexity &  Bandwidth cost & Latency & In-cache operations & Parallelism \\
     \toprule
     \multirow{2}{*}{Expand} & \multirow{2}{*}{$O(\flop)$} &  reading $(b*(\nnz(\mA) + \nnz(\mB))) $ bytes & \multirow{2}{*}{Negligible} & \multirow{2}{*}{manipulating local bins} & cols of $\mA$  and \\
      &  &  writing $(b* \flop) $ bytes &  &  & rows of $\mB$  per thread \\
      \midrule
    Sort & $O(\flop)$ &  reading $(b*\flop)$ bytes & Negligible & shuffling $(4 * b*\flop)$ bytes & bins per thread \\
    \midrule
     Compress & $O(\flop)$ &  writing $(b*\nnz(\mC))$ bytes & Negligible & read/write $(b*\flop)$ bytes & bins per thread \\
    \bottomrule

    \end{tabular}
     \\
     
    \caption{Computational complexity and data access patterns in different phases of PB-SpGEMM.}
   
    \label{tab:complxity}
\end{table*}

{\bf In-cache sorting.}
Since we sort $\hat{\mC}$ containing $\flop$ tuples, four passes over the data directly from memory can be the performance bottleneck.
Fortunately, bins can help in this case if tuples in a bin fit in L3 or L2 cache.
In many practical problems, we can indeed fit a bin into L2 cache. 
For example, consider an ER matrix $\mA$ with 1M rows, 1M columns and 4M nonzeros. 
When computing $\mA^2$, we generate 16M tuples in expectation.
If 1K bins are used, each bin will contain 16K tuples. 
If we use 4 byte keys (as described in the previous paragraph) and 8 byte payloads, we need 192KB memory to store all tuples in a bin.
This can easily fit in L2 cache of most modern processors. 
Multiplications with high compression ratios and matrices with denser rows can create problems for some bins as tuples may not fit in L2 cache. 
In these cases, we either use more bins or create bins with variable ranges of rows. 
Hence, our algorithm reads a bin from memory and perform radix sort on data stored in cache.
The sorted data can then be compressed while it is still in the cache.
Hence, the sorting phase reads $(b*\flop)$ bytes.

%{\bf Complexity.}
%Computation complexity. Memory bandwidth cost: reading $(b*\flop)$ bytes from memory. Cache

%In our preliminary evaluation, we compare the performance against some most popular algorithms like the std::sort, vergesort, timsort, qsort, pdqsort, the radix sort outperforms all of them.

\subsection{Compression} 
After we sort each bin, tuples with the same (rowid, colid) pair are stored in adjacent locations.
Then, in the compression phase, we sum numeric values from tuples with the same (rowid, colid).
As shown in Algorithm~\ref{algo:pb-spgemm}, compression can be performed independently in each bin because bins do not share tuples with the same rowid. 
Hence, a thread can compress tuples in a bin sequentially, while other threads compress other bins in parallel.

As tuples are already sorted within a bin, compression is done by scanning the tuples in the sorted order.
We implement this using two in-place pointers, which only walk the array once. 
The first pointer $(p_1)$ walks through the array, the second pointer $(p_2)$ maintains the location to be merged. 
Every time when $p_1$ points to a new location it checks with $p_2$, if the keys of the two tuples are the same, simply add the numeric value of the first tuple to the second, if not, we move $p_2$ to the next location and copy the tuple2 there, keep doing this until the $p_1$ reach the end of the array.

Table~\ref{tab:data_access} summarizes the computational complexity, bandwidth and latency costs, in-cache operations and parallelism schemes for all phases of PB-SpGEMM.
\section{Experiment Setup}

\subsection{Software}
We evaluate the performance of PB-SpGEMM against some state-of-the-art column SpGEMM algorithms, namely HeapSpGEMM, HashSpGEMM and HashVecSpGEMM.
All of these algorithms have memory access patterns of a typical column SpGEMM algorithm discussed earlier. 

\begin{itemize}[leftmargin=*]
%\item \textbf{PB-SpGEMM} implements the outer-product based SpGEMM described in this paper, which saturates memory band-width by using the propagation blocking technique and performs in-cache sorting and merging to generate the final output.
\item \textbf{HeapSpGEMM} is a column SpGEMM algorithm that uses heaps to merge  columns~\cite{azad2016exploiting}. To multiply two $n\times n$ ER matrices with $d$ average nonzeros per column, HeapSpGEMM complexity is $O(nd^2\log{d})$, 
where the $\log{d}$ term comes from manipulating heaps. Hence, HeapSpGEMM can be efficient for matrices with small $d$, but can be expensive for relatively dense matrices.
Each column of $\mC$ can be formed in parallel using thread-private heaps. 
%is space efficient because it only requires $O(nnz(a_i))$ memory to accumulate each row of the output.

\item \textbf{HashSpGEMM} uses a hash table to merge columns\cite{nagasaka2017high, nagasaka2019performance}.  
Its complexity is $O(nd^2)$ for ER matrices assuming that hash tables have limited collisions. 
Each column of $\mC$ can be formed in parallel using thread-private hash tables.
\item \textbf{HashVecSpGEMM} is a variant of hash algorithm, which utilizes vector registers for hash probing~\cite{ nagasaka2019performance}. 
HashVecSpGEMM may preform better when the collision in the hash table is high.
\end{itemize}

Previous work \cite{nagasaka2019performance} has shown that the optimized heap and hash algorithms largely outperform Intel MKL and Kokkos-kernel. 
Considering this, we will not include those algorithms and software in our evaluation.
All implementations are compiled by GCC-8.2.0, with flags "-fopenmp" "-O3" "-m64" "-march=native" enabled.

\begin{table}[!t]
\begin{center}

\begin{tabular}{lll}
\hline
                     & \textbf{Skylake-SP}      & \textbf{POWER9}  \\ \hline
%\textbf{CPU}         &                          &                  \\ \hline
CPU Model           & Intel Xeon Platinum 8160 & IBM POWER9       \\
Architecture         & x86\_64                  & ppc64le          \\
\#Sockets            & 2                        & 2                \\
\#Cores/socket       & 24                       & 20               \\
Clock                & 2.1GHz                   & 3.8GHz           \\
L2 cache             & 1024KB/core              & 512KB/two cores  \\
L3 cache             & 33792KB/socket           & 10240KB/two cores\\ \hline
%\textbf{Memory}      &                          &                  \\ \hline
Memory Size                 & 250GB                    & 1TB              \\
Bandwidth            & 100GB/s                  & 250GB/s          \\ \hline
%\textbf{Environment} &                          &                  \\ \hline
%OS                   & Red Hat 7.6 (Maipo)      & Ubuntu 18.04 LTS \\ \hline
\end{tabular}

\caption{The configuration of hardware used for evaluation}
\label{tab:hardware}

\end{center}
\end{table}

\begin{table}[!t]
\begin{center}

\begin{tabular}{@{}lllll@{}}
\toprule
              & Copy    & Scale   & Add      & Triad    \\ \midrule
single socket & 47.40 & 46.85 & 54.00  & 57.04  \\
dual socket   & 97.73 & 87.43 & 107.00 & 108.42 \\ \bottomrule
\end{tabular}
\caption{Stream benchmark result of the evaluation platform}

\label{tab:ch4-stream}

\end{center}
\end{table}

\subsection{Hardware}
In our experiments, we use a dual-socket Intel Skylake system and an IBM POWER9 system as described in Table \ref{tab:hardware}. 
Since most of our experiments are conducted on the Skylake processor, we examine its memory system carefully using the STREAM benchmark \cite{McCalpin2007}.
Table \ref{tab:ch4-stream} shows the sustainable memory bandwidth for Copy, Scale, Add and Triad benchmarks on single and dual sockets of the Skylake system. 
Hence, we expect that PB-SpGEMM should attain $\sim 55$ GB/s on a single socket and  $\sim 105$ GB/s on two sockets of Skylake. 
While PB-SpGEMM indeed attains $\sim 55$ GB/s in every phase, its dual socket performance falls short of STREAM benchmark. 
Hence, we primarily focus the single-socket performance because memory bandwidth is harder to predict in Non-Uniform Memory Access (NUMA) domains. 
We also show some results with both sockets and explain dual socket performance in Sec.~\ref{sec:multi-socket}.
When experimenting with a single socket, we set "OMP\_PLACES" to "cores", "OMP\_PROC\_BIND" to "close", and restrict the memory allocation to NUMA node 0 by using "numactl --membind=0".

\subsection{Dataset}
\label{sec:dataset}
We closely follow a recent paper~\cite{nagasaka2019performance} to select matrices to test SpGEMM algorithms. 
We use the R-MAT recursive matrix generator to generate synthetic matrices.
RMAT has four parameters $a,b,c, $ and $d$. 
ER matrices are generated with  a=b=c=d=0.25, and Graph-500 matrices are generated with a=0.57, b=c=0.19, d=0.05.
Here, we refer to the latter graphs as RMAT. 
For random graphs, edge factor denotes the average number of nonzeros in a row or column.
A matrix of scale $k$ has $2^k$ rows and columns.
We also use 12 real-world matrices from SuiteSparse Matrix Collection ~\cite{florida} as shown in Table \ref{tab:real-world-graphs-summary}.
In our experiments, we randomly generate two random matrices and multiply them or multiply a real matrix with itself. 
These computations cover some representative applications such as Markov clustering~\cite{HipMCL} and triangle counting~\cite{triangles}. 
Due to space limitation, we did not explore other application scenarios such as multiplying a square matrix by a tall-and-skinny matrix as needed in betweenness centrality algorithms.

\begin{table}[!t]
\begin{center}

\begin{adjustbox}{max width=\linewidth}
\begin{tabular}{|lllrrrr|}
\toprule

Graph & \multicolumn{1}{r|}{\textbf{n(A)}} & \multicolumn{1}{r|}{nnz(A)} & \multicolumn{1}{r|}{d(A)} & \multicolumn{1}{r|}{flops} & \multicolumn{1}{r|}{nnz(C)} & cf \\ \hline
2cubes\_sphere & 101.5K & 1.6M   & 16.23 & 27.5M  & 9.0M  & 3.06  \\ \hline
amazon0505     & 410.2K & 3.4M   & 8.18  & 31.9M  & 16.1M & 1.98  \\ \hline
cage12         & 130.2K & 2.0M   & 15.61 & 34.6M  & 15.2M & 2.14  \\ \hline
cant           & 62.5K  & 4.0M   & 64.17 & 269.5M & 17.4M & 15.45 \\ \hline
hood           & 220.5K & 9.9M   & 44.87 & 562.0M & 34.2M & 16.41 \\ \hline
m133\_b3       & 200.2K & 800.8K & 4.00  & 3.2M   & 3.2M  & 1.01  \\ \hline
majorbasis     & 160.0K & 1.8M   & 10.94 & 19.2M  & 8.2M  & 2.33  \\ \hline
mc2depi        & 525.8K & 2.1M   & 3.99  & 8.4M   & 5.2M  & 1.6   \\ \hline
offshore       & 259.8K & 4.2M   & 16.33 & 71.3M  & 69.8M & 3.05  \\ \hline
patents\_main  & 240.5K & 560.9K & 2.33  & 2.6M   & 2.3M  & 1.14  \\ \hline
scircuit       & 171.0K & 958.9K & 5.61  & 8.7M   & 5.2M  & 1.66  \\ \hline
web-Google     & 916.4K & 5.1M   & 5.57  & 60.7M  & 29.7M & 2.04  \\ \hline
\end{tabular}
\end{adjustbox}

\end{center}
\caption{A list of real-world graphs used in our evaluation}
\label{tab:real-world-graphs-summary}
\end{table}

\begin{figure}[!t]
\centering
\begin{subfigure}{0.48\linewidth}
  \centering
  \includegraphics[width=\linewidth]{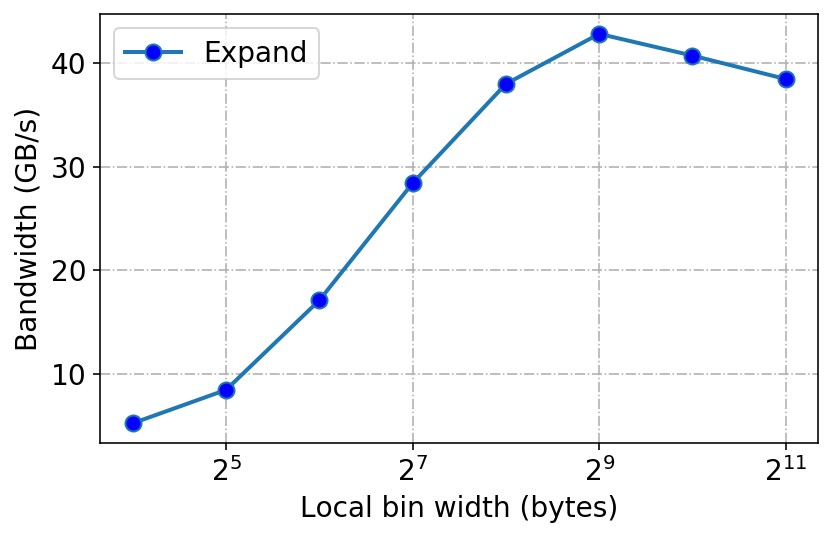}

  \caption{Local bin width}
  \label{fig:PB-SpGEMM-parameters-a}
\end{subfigure}
\begin{subfigure}{0.48\linewidth}
  \centering
  \includegraphics[width=\linewidth]{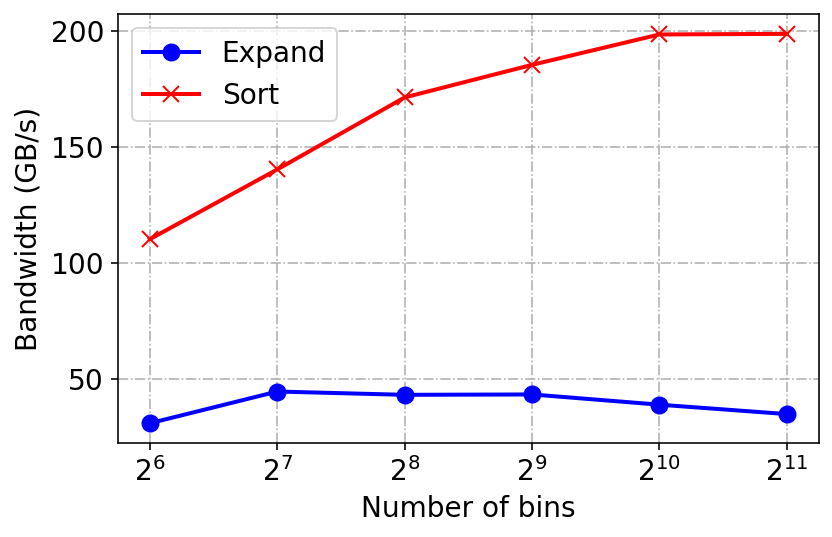}
  
  \caption{The number of bins}
  \label{fig:PB-SpGEMM-parameters-b}
\end{subfigure}

\caption{Impact of bin width and the number of bins, data based on ER matrices with scale 20 and edge factor 4}

\label{fig:PB-SpGEMM-parameters}
\end{figure}

\begin{figure*}[!t]
\centering
\begin{subfigure}{.65\textwidth}
  \centering
  \includegraphics[width=\linewidth]{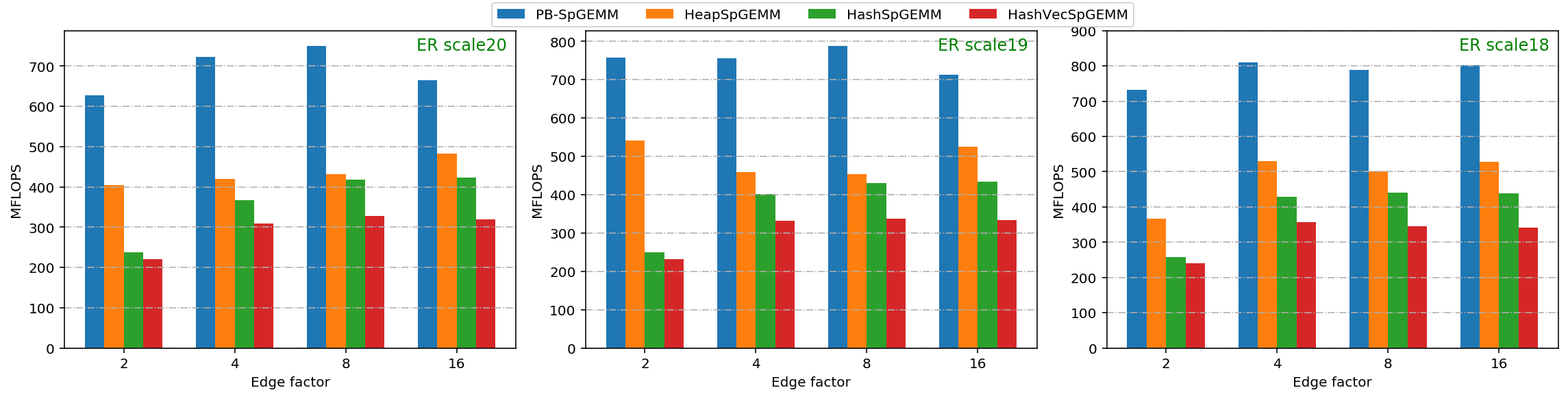}
  \vspace{-15pt}
  \caption{Performance scaling with matrix density}
  \label{perf-ER}
\end{subfigure}
\begin{subfigure}{.24\textwidth}
  \centering
  \includegraphics[width=\linewidth]{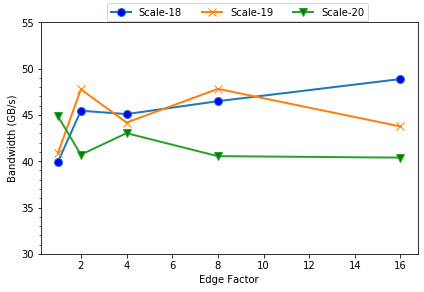}

  \caption{Sustained bandwidth}
  \label{bw-ER}
\end{subfigure}%

\caption{Performance and bandwidth evaluation with ER matrices on a single socket (24 cores) from Skylake}
\label{evaluation-ER}
\end{figure*}

\begin{figure*}[!t]
\centering
  \includegraphics[width=0.8\linewidth]{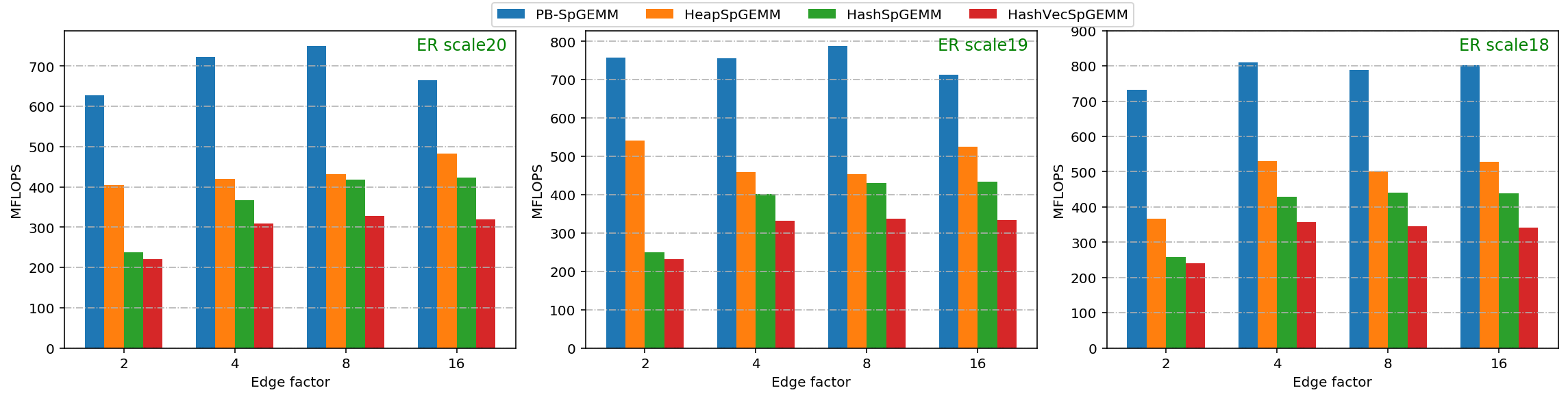}

 \caption{Performance evaluation with ER matrices on a single socket (20 cores) from Power9}
\label{fig:evauation-er-p9}
\end{figure*}

\begin{figure*}[!t]
\centering
\begin{subfigure}{.65\textwidth}
  \centering
  \includegraphics[width=\linewidth]{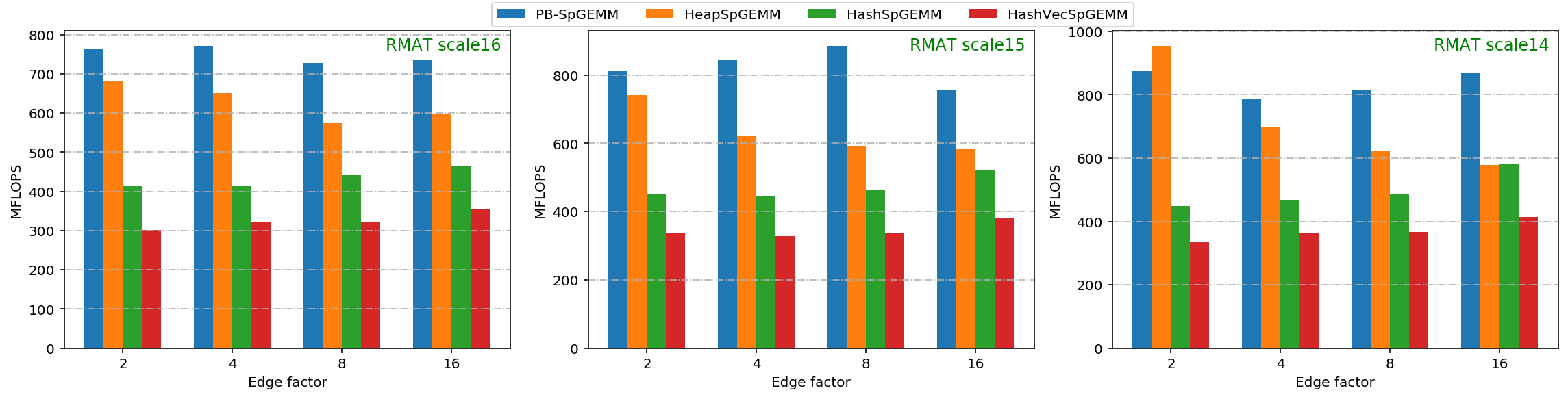}

  \caption{Performance scaling with matrix density}
  \label{perf-RMAT}
\end{subfigure}
\begin{subfigure}{.24\textwidth}
  \centering
  \includegraphics[width=\linewidth]{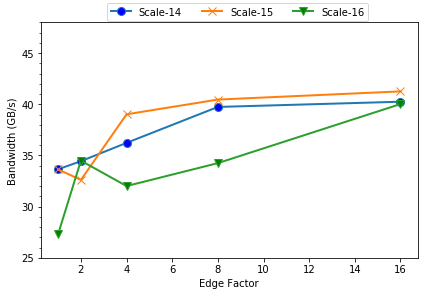}

  \caption{Sustainable bandwidth}
  \label{bw-RMAT}
\end{subfigure}%

\caption{Performance and bandwidth evaluation with RMAT matrices on a single socket (24 cores) from Skylake}
\label{fig:evaluation-rmat}
\end{figure*}

\begin{figure*}[!t]
\centering
  \includegraphics[width=0.8\linewidth]{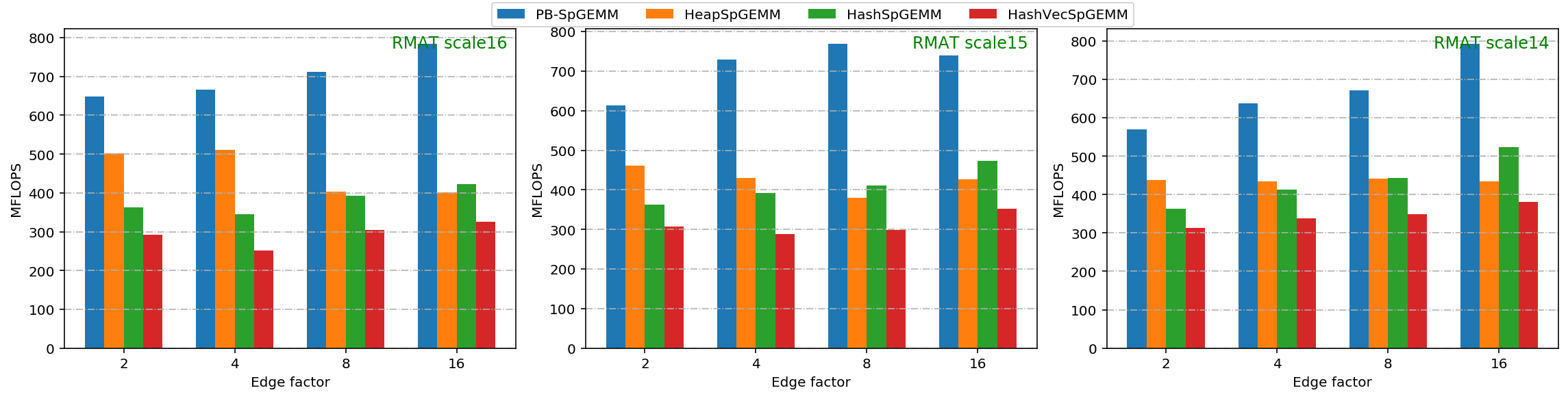}

 \caption{Performance evaluation with RMAT matrices on a single socket (20 cores) from Power9}
\label{fig:evauation-rmat-p9}
\end{figure*}

\section{Results}

\subsection{Select parameters of PB-SpGEMM}

\begin{figure*}[!t]
    \centering
    \includegraphics[width=0.8\linewidth]{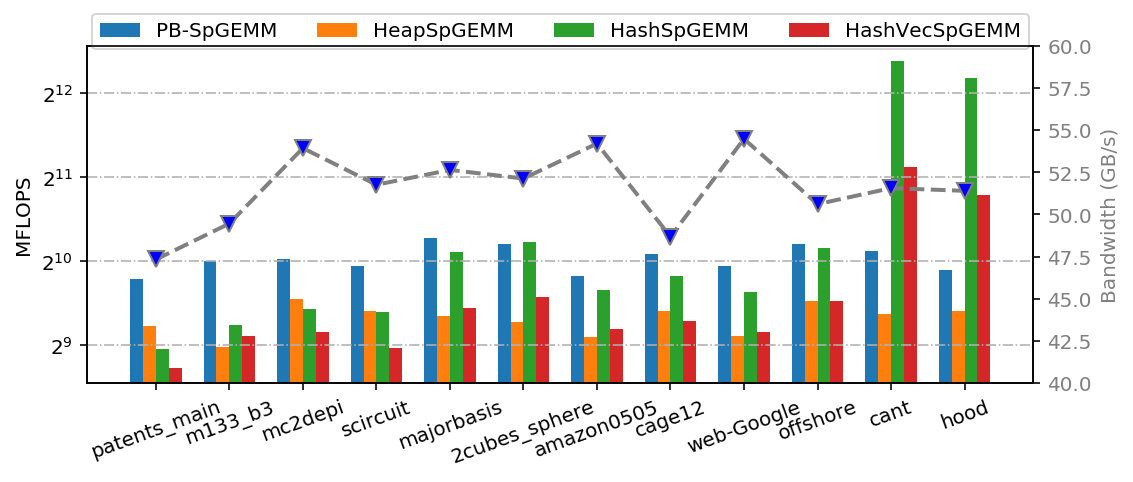}
    
    \caption{Performance evaluation with real matrices (matrix squaring) on a single socket of the Skylake server. From left to right, we sort matrices in the ascending order of the compression factor}
    \label{fig:perf-real}
\end{figure*}

In the expand phase, local bins are used to improve data locality, the propagation of tuples will be blocked into small bins, and they are moved to global shared memory. A local bin should be large enough to have good utilization of a cache line so that it can be send to global bins without wasting memory bandwidth.
Furthermore, the total size of local bins per thread should be smaller than the L2 cache~\footnote{For system that has multiple threads per core, it should be counted by core.}. 
Hence, number of bins and local bin width are two parameters upon which the performance of  PB-SpGEMM depends.  
Fig.~\ref{fig:PB-SpGEMM-parameters-a} shows our experiment where we vary the width of local bins to observe its impact on the performance of the expand phase. 
Smaller local bins do not utilize full cache lines, resulting in reduced sustained bandwidth. 
Based on this experiment, we used 512 bytes for every thread-private bin in all experiments in the paper.

The number of bins ($nbins$), on the other hand, is a tradeoff between the expand and sort phases.
Recall that sorting is performed independently in each bin.
Hence, increasing the number of bins ensures that tuples in a global bin fit in the cache.
However, increasing the number of bins also reduces the average bin size, which may reduce the bandwidth utilization of the expand phase.
Fig.~\ref{fig:PB-SpGEMM-parameters-b} shows the impact of $nbins$ over the expand and sorting phases. 
With more bins, radix sort can be entirely performed in L2 cache. 
Hence, in-cache sorting bandwidth can be as high as 200 GB/s.  
%Similar to Fig.~\ref{fig:PB-SpGEMM-parameters-b},
Hence, the number of bins is determined by L2/L3 cache size and total number of flop, and for most practical matrices, we use 1K or 2K bins.

\subsection{Overall performance of PB-SpGEMM with respect to the state-of-the-art}
At first we discuss the overall performance of PB-SpGEMM with respect to state-of-the-art column SpGEMM algorithms. 
Here, we consider both ER, RMAT, and real matrices as discussed in Sec~\ref{sec:dataset}.

{\bf Performance with ER random matrices.}
Fig.~\ref{perf-ER} compares the performance of PB-SpGEMM, HeapSpGEMM, HashSpGEMM, and HashVec-SpGEMM with ER matrices of various scales and edge factors on a single socket of the Skylake server. 
Here, we multiply two ER matrices with the same scale and edge factor. 
For a given scale, the performance of PB-SpGEMM is stable (between 700 and 800 GFLOPS) and is better than column SpGEMM algorithms for all edge factors considered.
This performance of PB-SpGEMM can be explained by Fig.~\ref{bw-ER} that reports the sustained bandwidth of PB-SpGEMM.
We observed that the sustained bandwidth on a single socket is between 40 and 50 GB/s, which are close to the STREAM benchmark.
Since RE matrices have $\cf=1$, the lower bound of AI is $\frac{1}{80}$ flops/byte according to Eq.~\ref{AI_outer}.
Hence, the performance of PB-SpGEMM should be at least $40*\frac{1}{80} = 500$ MFLOPS when the sustained bandwidth is 40 GB/s and  at least $50*\frac{1}{80} = 625$ MFLOPS when the sustained bandwidth is 50 GB/s.
Fig.~\ref{perf-ER} confirms that PB-SpGEMM's performance remains close to this lower bound estimate.
By comparison, hash and heap algorithms have lower performance primarily because of their irregular memory accesses.
However, as we increase the edge factor, the performance of column SpGEMM may increase because of increased utilization of cache lines.

Similar performance is observed on the Power9 system as shown in Fig.~\ref{fig:evauation-er-p9}.
As observed on Skylake, PB-SpGEMM performs better than column SpGEMM algorithms and its performance remains relatively stable for various matrix size and sparsity.

{\bf Performance with RMAT random matrices.}
Fig.~\ref{perf-RMAT} compares the performance of SpGEMM algorithms with RMAT matrices of various scales and edge factors on a single socket of the Skylake server. 
Here, we multiply two RMAT matrices with the same scale and edge factor. 
As with the ER matrices, the performance of PB-SpGEMM remains between 700 and 900 GFLOPS and is generally better than column SpGEMM algorithms.
However, Fig.~\ref{bw-RMAT} shows that PB-SpGEMM attains lower sustained bandwidth between 30 and 40 GB/s.
The reason behind this lower than STREAM bandwidth is the skewed degree distributions in RMAT matrices, resulted in variable-size bins. 
Load imbalance in different bins makes the expansion phase less bandwidth efficient. We will discuss more in the scalability results.

Similar performance is observed on the Power9 system as shown in Fig.~\ref{fig:evauation-rmat-p9}.
As observed on Skylake, PB-SpGEMM performs better than column SpGEMM algorithms and its performance remains relatively stable for various matrix size and sparsity. 

\begin{figure}[!t]
  \centering
  \includegraphics[width=0.95\linewidth]{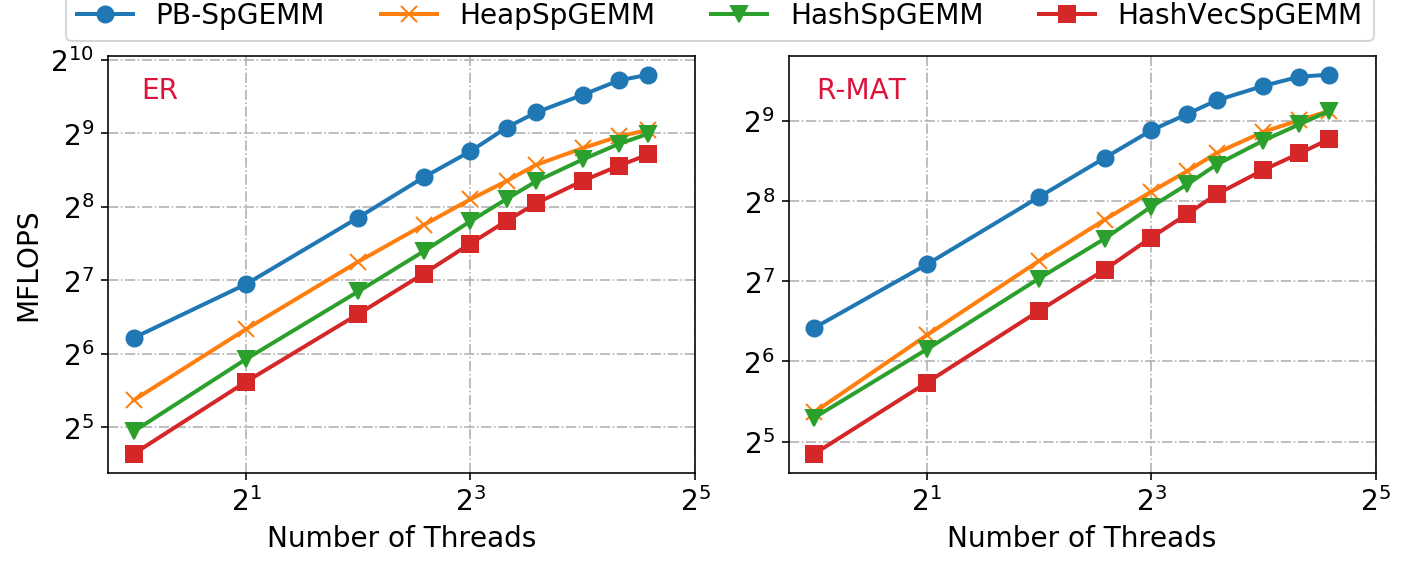}
 
\caption{Scalability on ER (left) and R-MAT matrices (right). Both are scale 16 matrices with edge factor of 16.}

\label{fig:scalability}
\end{figure}

\begin{figure}[!t]
  \centering
  \includegraphics[width=0.95\linewidth]{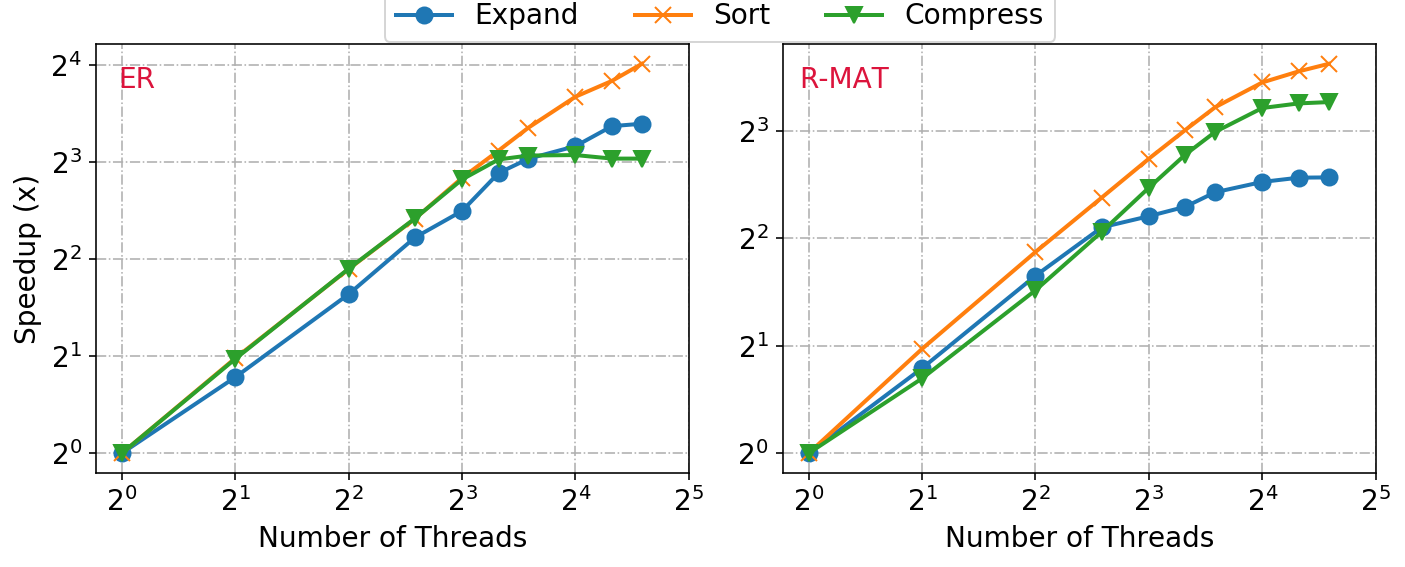}

\caption{Scalability breakdown on ER (left) and R-MAT matrices (right). Both are scale 16 matrices with edge factor 16.}

\label{fig:scalability-breakdown}
\end{figure}

{\bf Performance with real matrices.}
Fig.~\ref{fig:perf-real} shows the performance of SpGEMM algorithms when squaring real matrices on a single socket of the Skylake server.  
As before, the sustained bandwidth of PB-SpGEMM is between 47 and 55 GB/s and its performance is relatively stable.
In Fig.~\ref{fig:perf-real}, we sort matrices in the ascending order of the compression factor (from left to right).
PB-SpGEMM is generally faster than its peers.

\subsection{Scalability}
Fig.~\ref{fig:scalability} shows the strong scaling  of SpGEMM algorithms from 1 to 24 threads within a socket of the Skylake processor. 
We observe that PB-SpGEMM runs faster that column SpGEMM on all concurrencies.
All SpGEMM algorithms scale well within a socket. 
On 24 cores, PB-SpGEMM attains about $16\times$ speedup for ER matrices and $10\times$ speedup for RMAT matrices.
For RMAT matrices, PB-SpGEMM does not scale well on high thread counts because of the load imbalance caused by highly skewed nonzero and $\flop$ distributions. 
We tried to eliminate the load imbalance by variable length bins, but this can lead to lower sustained bandwidth as was observed in Fig.~\ref{bw-RMAT}.

% \subsection{Understanding the performance of PB-SpGEMM}

% \begin{figure}[h]
%   \centering
%   \includegraphics[width=0.95\linewidth]{figures/percentage.png}
% \caption{Execution time breakdown}
% \label{fig:excution-breakdown}
% \end{figure}

% \begin{figure*}[!t]
%     \centering
%     \includegraphics[width=0.8\linewidth]{figures/perf-real-by-cf.png}
%     \caption{Performance evaluation on real matrices with respect to compression ratio}
%     \label{fig:perf-real-by-cr}
% \end{figure*}

\begin{figure}[!t]
    \centering
    \includegraphics[width=0.95\linewidth]{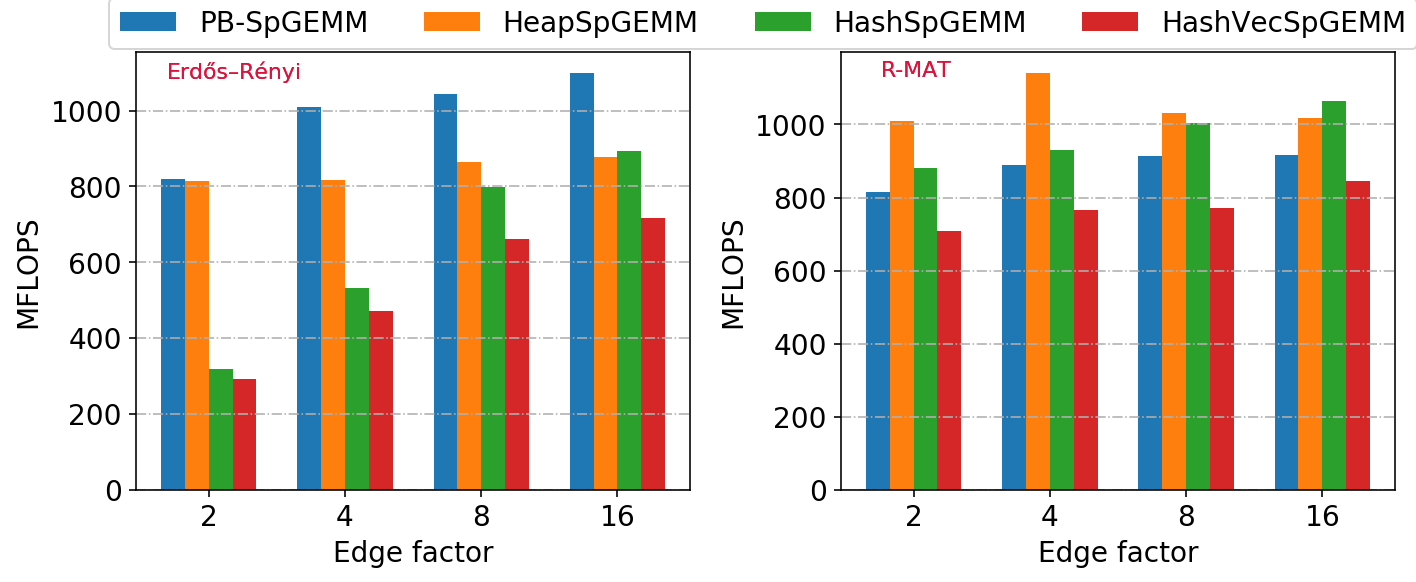}

    \caption{Multi-socket performance of SpGEMM algorithms on Skylake system}
 
    \label{fig:perf-multisocket}
\end{figure}

\subsection{Dual-socket Performance}
\label{sec:multi-socket}
Thus far, we have only considered the performance of SpGEMM algorithms on a single socket of Skylake and Power9 processors. 
Fig.~\ref{fig:perf-multisocket} shows the performance of SpGEMM algorithms on dual socket Skylake processor. 
PB-SpGEMM still runs faster for ER matrices, but runs slightly slower than heap algorithm for RMAT matrices. 
This lower-than-expected performance of PB-SpGEMM on NUMA systems is due to inter-socket communication contentions.
If a bin is allocated on socket-1 in the expand phase and sorted by a thread from socket-2, the performance of PB-SpGEMM depends on cross-socket memory bandwidth. 
We checked cross-socket memory bandwidth empirically by placing data in one socket and accessing from another socket in a STREAM copy kernel.
Table~\ref{tab:ch4-bw-latency} shows the local and remote access bandwidth and latency. 
Memory latency was measured by Intel Memory Latency Checker.
We observe that cross-socket access is much slower than local access.
Hence, PB-SpGEMM's performance relies saturating the memory bandwidth, it is affected by lower cross-socket bandwidth. 
Note that column SpGEMM algorithms are not significantly affected by cross-socket bandwidth because they generate one column at a time, where the active column usually stays in cache. 

In the Master's thesis of the first author, we tried to improve the dual socket performance by partitioning $\mA$ into two $\frac{n}{2}\times n$ matrices and multiply each part with $\mB$ independently in two sockets. 
This partitioned PB-SpGEMM partially mitigates the cross-socket bandwidth problem, but it does not perform uniformly well for all matrices due to the additional cost of reading $\mB$ more than once. 
We did not cite the thesis as per the double blind policy.

\begin{table}[!t]
\begin{center}

\begin{tabular}{@{}l|l|l@{}}
\toprule
                   & NUMA socket 0 & NUMA socket 1 \\ \midrule
NUMA socket 0  & 50.26GB/s and  88.1ns       & 33.36GB/s and 147.4ns      \\ 
NUMA socket 1  & 34.06GB/s and 146.7ns       & 50.12GB/s and 88.3ns       \\ \bottomrule
\end{tabular}

\caption{NUMA local and cross-socket memory bandwidth and latency on Skylake}

\label{tab:ch4-bw-latency}

\end{center}
\end{table}

%However, this could be more complicated if we consider a multi-socket system where Non-Uniform Memory Access (NUMA) \cite{Lameter:2013:NO:2508834.2513149} is enabled, local memory latencies and cross-socket memory bandwidth and memory latencies will vary significantly. Table \ref{tab:ch4-bw-latency} describes the local access and cross-socket access memory bandwidth from STREAM copy kernel, as well as the memory latency stats that reported by Intel Memory Latency Checker.

% \begin{figure*}[!t]
% \centering
% \begin{subfigure}{0.8\linewidth}
%   \centering
%   \includegraphics[width=\linewidth]{figures/scalability.png}
%   \caption{The overall performance scaling with increasing number of threads}
%   \label{fig:scalability-overall}
% \end{subfigure}
% \begin{subfigure}{0.8\linewidth}
%   \centering
%   \includegraphics[width=\linewidth]{figures/scalability-breakdown.png}
%   \caption{PB-SpGEMM speedup breakdown, all phases demonstrate decent speedup until hit memory bandwidth bound, sorting gets further improvement due to cache utilization}
%   \label{fig:scalability-breakdown}
% \end{subfigure}%
% \caption{Scalability on ER (left) and R-MAT matrices (right), both are scale 16 edge and factor 16}
% \label{fig:scalability}
% \end{figure*}

\section{Conclusions}
With the rise of sparse and irregular data, SpGEMM has emerged as an important operation in many scientific domains.  
Over the past decade, the state-of-the-art of parallel SpGEMM algorithms has progressed significantly. 
However, understanding the performance of SpGEMM algorithms remains a challenge without an established performance model.
Relying on the fact that SpGEMM is a bandwidth-bound operation, we used the Roofline model to develop bounds for SpGEMM algorithms based on column-by-column merging and the expand-sort-compress strategy. 
We conclude the paper with the following key findings:
\begin{enumerate}[leftmargin=*]
    \item We can estimate the arithmetic intensity (AI) of an SpGEMM algorithm based on the compression factor of the multiplication and number of bytes needed to store each nonzero. 
    \item The attainable performance of an algorithm is $(AI*\beta)$, where $\beta$ is the memory bandwidth. This peak performance can only be attained if the algorithm saturates the bandwidth.
    We showed that existing column SpGEMM algorithms do not attain peak performance according to the Roofline model because of irregular data accesses and underutilization of cache lines. 
    \item We develop a new algorithm based on outer product of matrices. This algorithm called PB-SpGEMM uses propagation blocking to group multiplied tuples into bins and then sort and merge tuples independently in each bin.   
    \item PB-SpGEMM approximately saturates the memory bandwidth in all of its three phases and attains performance as predicted by the Roofline model.
    \item On a single socket, PB-SpGEMM performs better than the best column SpGEMM algorithms for multiplications with compression factors less than four.
    \item For multiplications with compression factors greater than four, HashSpGEMM is the best performer.
\end{enumerate}

\bibliographystyle{IEEEtran}
\bibliography{main}

\end{document}